# Unusual Polar Conditions in Solar Cycle 24 and their Implications for Cycle 25

Nat Gopalswamy[1], Seiji Yashiro[1,2], and Sachiko Akiyama[1,2]

[1]NASA Goddard Space Flight Center, [2]The Catholic University of America

ABSTRACT

We report on the prolonged solar-maximum conditions until late 2015 at the north-polar region of the Sun indicated by the occurrence of high-latitude prominence eruptions and microwave brightness temperature close to the quiet Sun level. These two aspects of solar activity indicate that the polarity reversal was completed by mid-2014 in the south and late 2015 in the north. . The microwave brightness in the south-polar region has increased to a level exceeding the level of cycle 23/24 minimum, but just started to increase in the north. The north-south asymmetry in the polarity reversal has switched from that in cycle 23. These observations lead us to the hypothesis that the onset of cycle 25 in the northern hemisphere is likely to be delayed with respect to that in the southern hemisphere. We find that the unusual condition in the north is a direct consequence of the arrival of poleward surges of opposite polarity from the active region belt. We also find that multiple rush-to-the-pole episodes were indicated by the prominence eruption locations that lined up at the boundary between opposite polarity surges. The high-latitude prominence eruptions occurred in the boundary between the incumbent polar flux and the insurgent flux of opposite polarity.

Key words: Sun: activity – Sun: filaments, prominences – Sun: radio radiation – Sun: magnetic fields

## 1. Introduction

One of the indicators of the solar maximum phase is the phenomenon of rush to the poles (RTTP; Altrock 2014) of polar crown prominences was recognized in the 19$^{th}$ century (see Cliver 2014 for a review). Babcock & Babcock (1955) speculated that the Sun's global field should reverse around solar maxima; Babcock (1959) confirmed it during the cycle-19 maximum. Hyder (1975) found the reversal to be synchronous with RTTP at each pole. RTTP results in polar crown filaments, which were suggested to be the source of high-latitude (HL) coronal mass ejections (CMEs) (Sheeley et al. 1980). The observational overlap between the Nobeyama Radioheliograph (NoRH; Nakajima et al. 1994) and the Large Angle and Spectrometric Coronagraph (LASCO; Brueckner et al. 1995) on board the Solar and Heliospheric Observatory (SOHO) confirmed the close association between polar prominence eruptions (PEs) and CMEs (Gopalswamy et al. 2003a, b). HL prominences indicate the presence of large-scale bipolar magnetic fields that need to disappear via the cancellation of the incumbent flux by the incoming surges before polarity reversal can occur. CMEs represent the process by which such regions are removed from the Sun (Low 1997) before the reversal.

In cycle 24, HL PEs (latitudes >60º) started appearing in of 2010 and 2012 in the north and south, respectively (Gopalswamy et al. 2012; Gopalswamy 2015). Using data from the Helioseismic and Magnetic Imager (HMI; Schou et al. 2012) and Atmospheric Imaging Assembly (AIA, Lemen et al. 2012) on board the Solar Dynamics Observatory (SDO), Karna et al. (2014) estimated the reversal epoch in the north as Mid-2012; Sun et al. (2015) estimated the epochs as November 2012 (north) and March 2014 (south). However, the subsequent evolution of the polar conditions was quite unexpected: zero polar magnetic field conditions prevailed for another three years as indicated by the lack of microwave brightness enhancement (MBE) and the continued HL PE activity in the northern hemisphere. The microwave observations provide information complementary to the polar field measurements. In this paper we revisit the issue of polarity reversal using the microwave signatures of polar activity.

## 2. Observations

We use 17-GHz microwave images obtained by the Nobeyama Radioheliograph (NoRH, Nakajima et al. 1994) that show PEs and MBEs. The brightness temperature of the quiet Sun at 17 GHz is ~10$^4$ K and MBE means exceeding this level. MBE in coronal holes (Kosugi et al. 1986; Gopalswamy et al. 1999a, 1999b; Nindos et al. 1999; Shibasaki et al. 2011; Prosovetsky & Myagkova 2011; Gopalswamy et al. 2012; Akiyama et al. 2013; Shibasaki 2013) is related to the enhanced magnetic fields in the upper chromosphere at the bottom of coronal holes (Gopalswamy et al. 1999a; Gopalswamy, Shibasaki & Salem 2000; Gopalswamy et al. 2012; Akiyama et al. 2013). MBEs are also observed from active regions (Shibasaki 2013; Selhorst et al. 2014). The microwave butterfly diagram (Gelfreikh et al. 2002; Gopalswamy et al. 2012; Shibasaki 2013) shows both kinds of MBEs. The butterfly diagram is a time latitude plot of microwave brightness temperature constructed from longitudinally-averaged NoRH synoptic charts (Shibasaki 1998).

The PE locations detected automatically from 17-GHz images (Gopalswamy et al. 2003a; Shimojo et al. 2006; Gopalswamy et al. 2012; Shimojo 2013; Gopalswamy 2015) clearly show solar activity variations such as RTTP. Since NoRH is ground based, only ~8 h of daily observations are possible. Since 2010, SDO/AIA provides continuous, high-cadence, and high-resolution images at 304 Å available at the Joint Science Operations Center (JSOC, http://jsoc.stanford.edu/data/aia/synoptic/). Automatic detection of PEs from these images has already been reported (Gopalswamy, Yashiro & Akiyama 2015). We use the Synoptic Optical Long-term Investigations of the Sun (SOLIS) magnetic data to study the poleward surges of unipolar flux (Hathaway 2015).

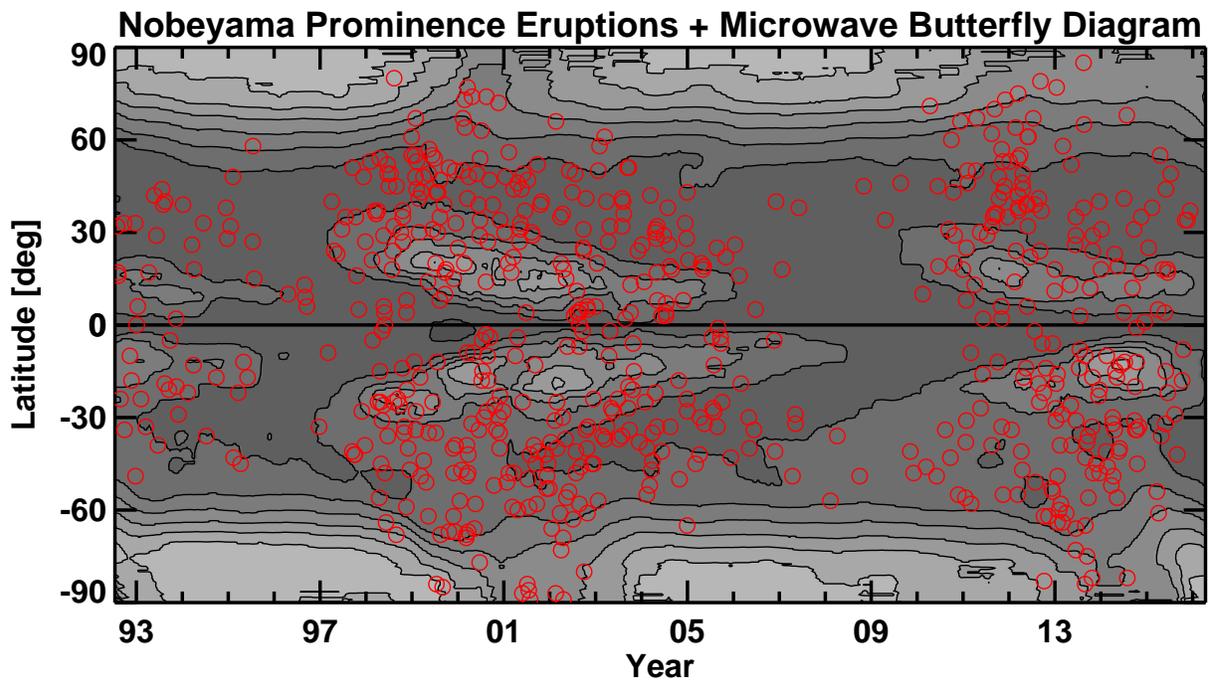

Figure 1. The locations (red circles) of NoRH microwave prominence eruptions (PEs) superposed on the most recent microwave butterfly diagram (contours and grey scale). A 13-rotation smoothing has been used along the time axis to eliminate the periodic variation due to solar B0-angle variation. The microwave butterfly diagram is constructed from NoRH synoptic charts made from daily-best microwave image take around local noon. The contour levels are at 10,000, 10,300, 10,609, 10,927, 11,255, 11,592, and 11,940 K. The HL MBE patches correspond to times of enhanced magnetic field strength. The LL MBE patches correspond to active regions.

The microwave butterfly diagram in Figure 1 shows two north-polar (1992-1999; 2002-2011) and three south-polar (1993-2001; 2003-2014; late 2015) MBE patches, which correspond to times of high polar magnetic field (B) strengths. The intervals between the MBE patches correspond to solar maxima. The most outstanding feature in Figure 1 is the lack of MBE in the north-polar region during 2012-2015 because B~0 (Gopalswamy et al. 2012). The three pairs of low-latitude (LL) MBEs correspond to active regions in cycles 22, 23, and 24.

PEs are very infrequent during minima (Figure 1). As activity rises, PEs occur at higher and higher latitudes, mimicking RTTP (Gopalswamy et al. 2003b; 2012). Most PEs are located at 30-50º latitudes except during maxima, when HL PEs occur. PEs are almost always associated with CMEs, and hence indicate sites of large-scale energy release (see e.g., Hori & Culhane 2002; Gopalswamy et al. 2003a; Gopalswamy, Yashiro & Akiyama 2015). Cessation of HL PEs roughly coincide with polarity reversals.

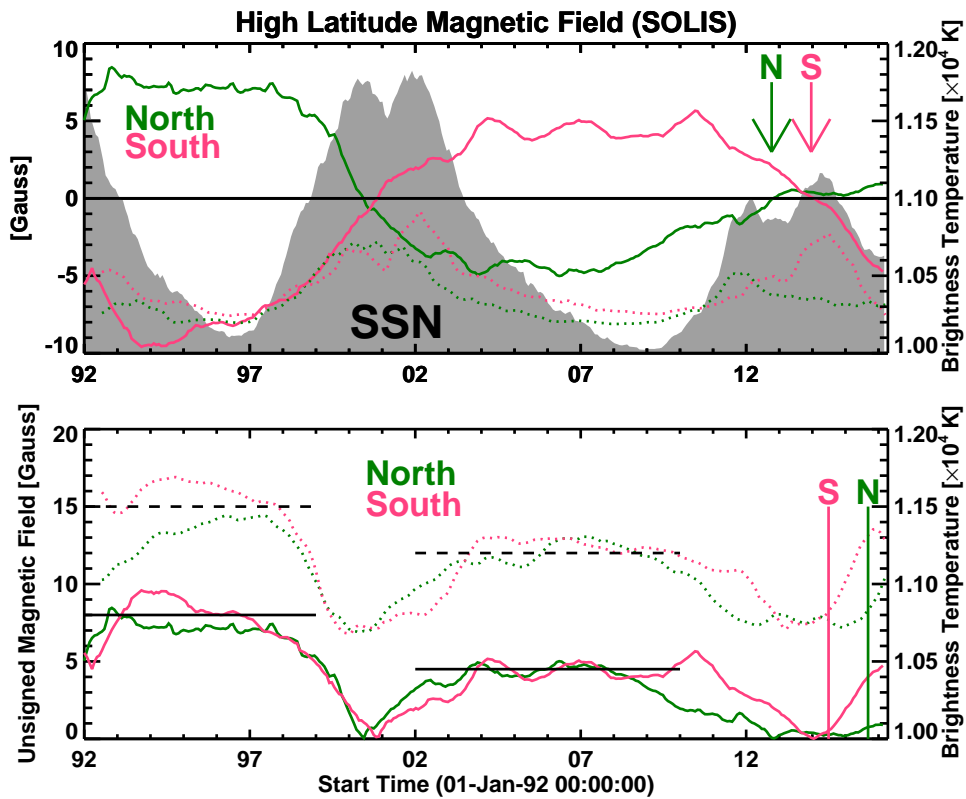

Figure 2. (Top) The polar B in the northern (green, solid line) and southern (pink, solid line) hemispheres averaged over latitudes poleward of 60º and over-plotted on the sunspot number (SSN, grey). The LL microwave Tb, averaged over 0-40º latitudes is also plotted in the two hemispheres (north: green dotted line, south: pink dotted line). Arrows N and S mark the start of B=0 times (north: October 2012; south: December 2013). (Bottom) HL Tb averaged over latitudes poleward of 60º (north: green dotted line, south: pink dotted line). Polar |B| is also plotted for reference (north: green solid line, south: pink solid line). The horizontal solid (dashed) lines indicate the drop in B (Tb) between the 22/23 and 23/24 minima. The vertical lines mark the times of reversal completion (north: N, September 2015; south: S, June 2014). The last B=0 time was during CR 2158 (2014 December 8 to 2015 January 4) in the north and a year earlier in the south (CR 2146 – 2014 January 15 to February 11). Note that the reversal occurs earlier in the south, which is different from cycle 23.

## 3. Analysis and Results

### 3.1 Polar Magnetic Field and Micowave Brightness

Figure 2 shows the time evolution B and the microwave brightness temperature (Tb) averaged over high (≥60º) and low (0-40º) latitudes in the two hemispheres along with the sunspot number (SSN). The two SSN peaks are also observed in LLTb, but with different contributions from the two hemispheres. In cycle 23, the northern hemipsheric Tb has a broad peak covering the two SSN peaks; the southern hemsipheric Tb has two distinct peaks, the second one being the largest. In cycle 24, the first Tb peak has equal contributions from the two hemispheres, while the second peak is pedominantly from the southern hemisphere (see also the contour plot in Figure 1).

The polar B becomes zero in the northern and southern hemipsheres around the first and second SSN (and Tb) peaks, respectivly. The zero-crossng pattern of B is clearly different in cycles 23 and 24: B~0 occurs in October 2012 in the north, but the reversal is complete only in September 2015, after a small B peak in early 2013. The first B~0 epoch in the north was identified with the reversal by Karna et al. (2014) and Sun et al. (2015). In the south, B~0 starts during late 2013, continues to become more negative, and exceeds the magnitude of the fluctuations in the north by June 2014 (reversal completion). The MBE information provides a key confirmation of this evolution. Our polar-B evolution is consistent with Petrie (2015) and Mordvinov et al. (2016); the small differences are due to different latitudinal and temporal averaging used.

The HL Tb evolution is highly asymmetric: in the north, Tb is hovering around the quiet-Sun values from the beginning of 2012 to mid-2015, but only briefly in the south (Figure 2). The steady increase in HL Tb occurs first in the south (June 2014) and then in the north (September 2015), consistent with the polar-B evolution. The different evolution of Tb in the north and south can also be clearly seen from the Tb contours in Figure 1. The south-polar Tb suddenly increases past the cycle-23 peak (Figure 2). Whether this high value will be sustained or not depends on the poleward transport of active region flux in the declining phase of cycle 24. In cycle 23, the rise in the new-cycle B started first in the north and then in the south, but happened within a year apart. Thus there is a clear change in north-south asymmetry in the reversal epoch between cycles 23 and 24.

The development of polar coronal holes is a key indication that the new-cycle flux is building up. The EUV synoptic chart in Figure 3 shows the deveopment of polar coronal during Carrington rotation 2174 (2016 February 18 to March 16). The synoptic chart was constructed from SDO/AIA 193 Å images available at http://jsoc.stanford.edu/data/aia/synoptic/. In the south, the polar coronal hole is well developed. In the north, the polar coronale hole can be seen only during only part of the rotation. This is the reason for a smaller, lonitudinall-averged MBE in the north polar region plotted in Figure 2.

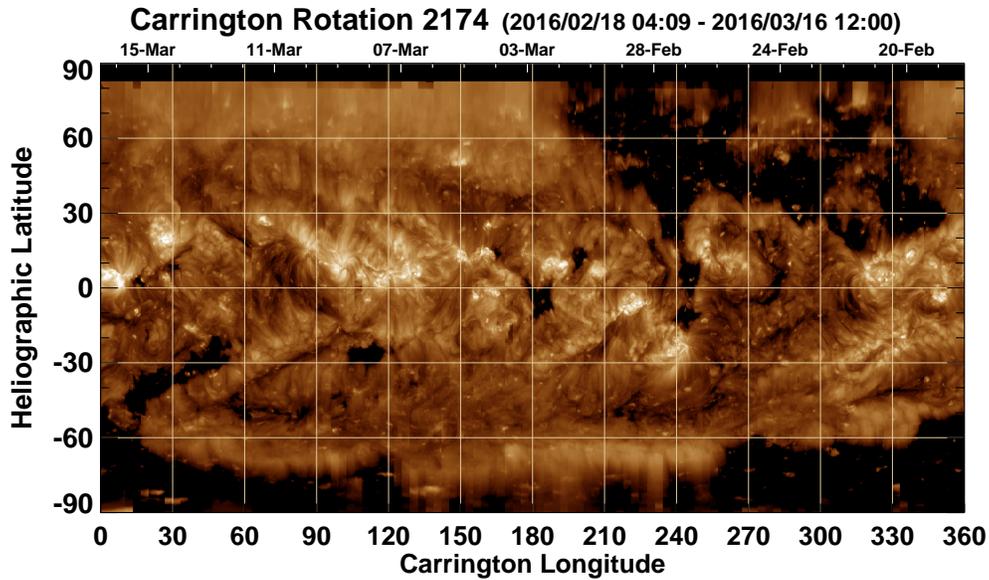

Figure 3. EUV synoptic chart for Carrington Rotation 2174 (2016 February 18 to March 16) constructed from SDO/AIA 193 Å images available at the at the Joint Science Operations Center (JSOC, http://jsoc.stanford.edu/data/aia/synoptic/). Central meridian strips from these images extracted every six hours are assembled as a synoptic chart. The resulting map has a spatial resolution of ~4 arcsec. The south polar coronal hole is fully formed. The north polar hole is observed only during part of the Carrrington rotation.

## 3.2 SDO Prominence Eruptions and Poleward Flux Surges

We have plotted the PE locations detected automatically from SDO 304 Å images on the magnetic butterfly diagram in Figure 4. Even though the PE locations were identified from all around the limb, the longitude information (whether east limb or west limb PE) is lost when plotted on the butterfly diagram because it is a time-latitude plot averaged over longitudes. The poleward flux surges are marked N1-N6 in the north and S1-S5 in the south. The surges are labled as in Sun et al. (2015) with additional surges N6 and S5. Mordvinov et al. (2016) combined N4 and N5 into a single surge (their label 13); N6 was not numbered; they also did not label S2 and S4. The PE locations lineup along boundaries of opposite polarity surges at latitudes below 60º, appearing as multiple RTTPs (Figure 4a). The HL PEs occur at the low-latitude boundary of the incumbent flux when the incoming surge is of opposite polarity (Figure 4b). The first set of HL PEs occur at the boundary between N1 (positive) and the incumbent flux (negative). There are no PEs between N2 and the polar flux because both are of the same polarity. N2 adds to the incumbent flux rather than canceling it and hence delays the reversal. N3 cancels some more incumbent flux, shrinking its area. The arrival N3 coincides with the reversal epoch identified by Sun et al. (2015). However, N4 and N5 thwart the reversal process because they are of the same polarity as the incumbent flux. Finally, N6 overwhelms the incumbent flux and permanently reverses the polarity by late 2015 (coinciding with the steady Tb increase in Figure 2). The last set of HL PEs were located at the boundary between the initial part of N6 and the incumbent flux. The reversal was complete only after the cessation of the HL PEs. Sun et al. (2015) used data until 2014 April 30, so the plume N6 was not included.

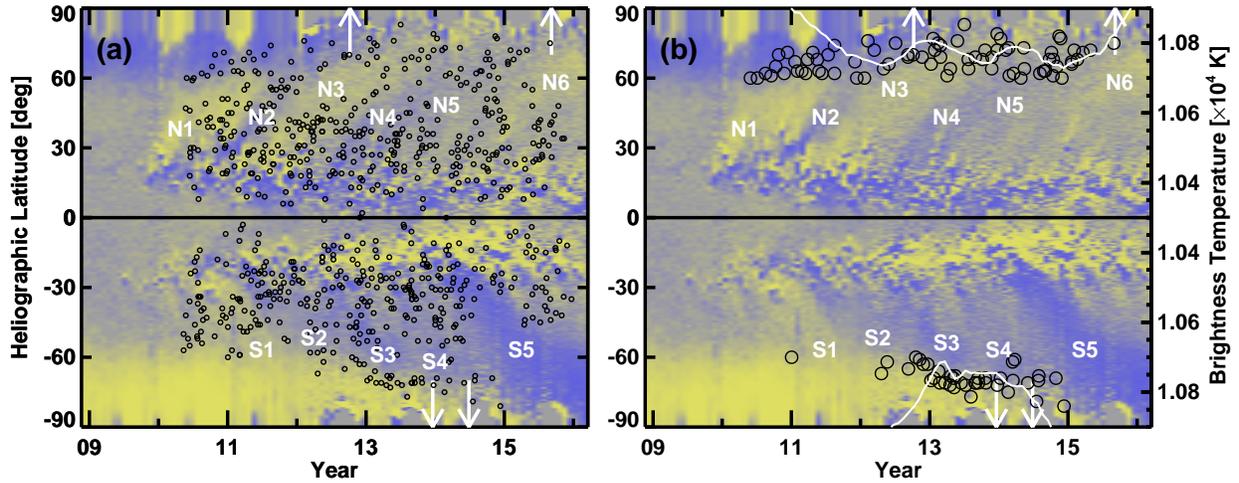

Figure 4. SDO/AIA 304 Å PE locations plotted on the magnetic butterfly diagram from SOLIS: (a) all PEs, (b) only HL PEs. HL Tb around the time of HL PEs (white curves) and the times of B~0 (first set of arrows) and reversal completion (second set of arrows) are marked. The poleward surges are labeled as N1-N6 in the north and S1-S5. The magnetic butterfly diagram was constructed from SOLIS synoptic charts and averaged over longitudes. The yearly artifacts in each hemisphere corresponds to times when the pole is not visible from Earth. SOLIS data gaps for a few rotations (2152-2155, 2163, 2166-2167) were filled by SDO/HMI data by calibrating between the two data sets. Note that low Tb values reached a minimum value of ~10,700 K three times during the interval of HL PE occurrence.

The evolution of the south polar flux is somewhat simpler. The surge S1 is positive, same polarity as the incumbent flux, so there are no HL PEs. S1 clearly delays the arrival of the maximum condition in the south polar region. Surges S2-S5 are negative, pressing against the inumbent positive flux and hence there is a steady chain of PEs. The strong S5 overwhelms any remaining incumbent flux, before establishing a steady unipolar negative flux (new magnetic cycle) in the middle of 2014. The cessations of southern HL PEs coincided with the steady increase in Tb.

Table 1. Various Time Marks during the Cycle-24 Maximum

|  | North | | South | |
| --- | --- | --- | --- | --- |
|  | Begin | End | Begin | End |
| Tb ~$10^4$ K | 04/08/2012 | 12/08/2014 | 03/21/2013 | 01/15/2014 |
| B ~0 | 10/08/2012 | 12/08/2014 | 12/19/2013 | 01/15/2014 |
| PE >60° | 08/07/2010 | 08/30/2015 | 04/20/2012 | 12/11/2014 |
| Reversal Completion | 09/07/2015 | | 06/28/2014 | |

Table 1 summarizes various time marks during the maximum phase of cycle 24 until the completion of the polarity reversal. The earliest is the appearance of PEs above 60º latitude in the north in August 2010 that continued for five years (the last PE was on 2015 August 30). In the south, HL PEs appeared from April 2012 to December 2014. Close-to-quiet-Sun Tb and B~0 conditions ended in December 2014 (north) and January 2014 (south). The sharp reversal in the south is clear because the B~0 condition lasts only for one Carrington rotation. The last instance of B~0 in Table 1 is in agreement with Sun et al. (2015) and Mordvinov et al. (2016) within a few months in the south. In the north, a similar agreement was found with Mordvinov et al. (2016), but not with Sun et al. (2015). Table 1 also shows that the times of reversal completion are close to the cessation times of HL PEs and several months after the end of B~0 conditions.

RTTP and the polarity reversal can be episodic (Howard &Labonte 1981; Topka et al. 1982; Makarov & Sivaraman 1989; Krivova & Solanki 2002; Gopalswamy et al. 2003; Ulrich & Tran 2013; Sun et al. 2015; Petrie 2015; Mordvinove et al. 2016), directly related to the series of poleward surges of alternating polarity. Surface flux transport simulations show that the emergence of a number of large active regions that violate Joy's law leads to the weak polar field during the cycle 23/24 minimum (Jiang et al. 2015). Emrgence of active regions violating Joy's law or Hale-Nicholson (HN) polarity rule is random, so solar-cycle prediction has inherent uncertainty. Yeates et al. (2015) investigated the origin of surge N2 (they labeld it as A in their figure 1) and showed that a set of two active regions violating Joy's law contributed significantly N2. Sources of other surges have been discussed in Sun et al. (2015) and Mordvinov et al. (2016).  N4 and N5 originated immdiately after the first SSN peak in 2012. In the southern hemipshere, S5 occurred around the time of the larger, second SSN peak. S5 had polarity opposite to that of the incumbent flux, resulting in the cancellation of all the incumbent flux and a quick buildup of the next-cycle polarity. This also led to the change in the north-south asymmetry in the phase of the reversal.

## 4. Discussion

The primary result of this paper is the prevalence of an unusually prolonged zero-field condition in the north-polar region as indicated by the continued occurrence of HL PEs and the lack of polar MBE. In the south, the zero-field condition is rather brief. The prolonged zero-field condition in the north is due to three poleward surges of the same (negative) polarity as the incumbent flux. We also found multiple RTTPs that correspond to the occurrence of PEs along boundaries of opposite-polarity surges. At high latitudes, PEs occurred along boundaries separating the incumbent flux and opposing insurgent flux. The high rate of HL PEs indicate that the flux cancellation in the polar region is an energetic process involving frequent energy release (PEs and CMEs).

The polar microwave brightness is a sensitive indicator of the polar field strength (Gopalswamy et al. 2012). The undulating behavior of the microwave brightness during the zero-field interval in the north is an accurate reflection of the alternating polarities of the incoming surges. While the surge N1 brings the Tb down to ~10,700 K (~quiet-Sun level), N2 increases it. N3 lowers the Tb again, while N4+N5 contribute to its increase. Finally, N6 brings the Tb down to the quiet-Sun level signaling the complete cancellation of the old flux; Tb soon increases as the new-polarity flux increases. There is no oscillatory behavior of Tb in the south because all surges after the first one are of the same polarity, opposing the incumbent flux. The polar flux steadily declines, reverses sign, and the new flux starts building up quickly in the south.

The extended zero-field interval in the north suggests that the onset of cycle 25 will be substantially delayed in the nothern hemisphere. The peak SSN of cycle 25 will depend on the level to which the polar field stength will be built during the declining phase of cycle 24. The south-polar B~0 condition is brief and the new-cycle flux has already increased to the peak levels of cycle 23, suggesting a norml cycle 25 in the southern hemisphere. The actual field strengths will depend on the poleward surge of trailing flux of active regions in the declining phase of cycle 24. If the weakness of cycle 24 continues in the declining phase, only weak poleward surges may result and hence a weaker cycle 25 is expected. These inferences are based on the expectation from the Babcock-Leighton model of solar cycles in that the polar field strength of one cycle determines the phase and amplitude of the next cycle (Babcock 1961; Leighton 1964; Wang & Sheeley 2009; Muñoz -Jaramillo et al. 2013; Cameron & Schüssler 2015; Jiang et al. 2015). Note that we are mainly talking about the cycle onset, while the precursor method usually predicts the maximum sunspot number in the next cycle and its epoch (Hathaway 2015, and references therein). The cross-correlation between polar and low-latitude microwave brightness temperatures do indicate a correlation with a lag of about half a solar cycle (Gopalswamy 2015).

## 5. Conclusions

The main conclusions of this paper are as follows:

1. The north polar region of the Sun had an unusually long stretch of near-zero magnetic field strength for more than three years. This was caused by surges of both polarities towards the north pole that prevented the buildup of the polar field until the end of 2015. The alternating surges caused the undulating pattern in the polar microwave brightness enhancement.

2. The continued occurrence of high-latitude prominence eruptions, the lack of microwave brightness enhancement, and the absence of polar coronal holes are consistent with the prolonged zero-filed condition in the north.

3. The end of zero-field condition is indicated by the steady increase in microwave brightness enhancement above the quiet-Sun level and the cessation of high latitude prominence eruptions.

4. In the southern hemisphere, most of the surges were of opposite polarity to that of the incumbent flux, so the reversal was completed in the middle of 2014. Because of the last intense surge, the south polar magnetic field rapidly increased as indicated by both SOLIS magnetic field data and microwave brightness enhancement.

5. We identified multiple rush to the poles episodes from PE locations. The PEs occurred at the boundary between poleward surges of opposite polarity. The high-latitude PEs occurred along the boundary separating the incumbent polar flux and the insurgent flux of opposite polarity.

6. There is a clear change in the north-south asymmetry of polarity reversal. For the past several cycles, north was reversing first. In cycle 24, the reversal was more than a year ahead in the southern hemisphere. The arrival of zero-field condition, however, was first in the north.

**Acknowledgments.** NoRH is currently operated by the Nagoya University in cooperation with the International Consortium for the Continued Operation of the Nobeyama Radio heliograph (ICCON). This work utilizes SOLIS data obtained by the NSO Integrated Synoptic Program (NISP), managed by the National Solar Observatory, which is operated by the Association of Universities for Research in Astronomy (AURA), Inc. under a cooperative agreement with the National Science Foundation. This work benefited from NASA's open data policy in using SDO data. Work supported by NASA's LWS TR&T program. We thank the anonymous referee for the helpful comments.